\documentclass[aps, pre, preprint, showpacs, footinbib]{revtex4-1}

\usepackage{natbib}
\usepackage{amssymb,amsmath, amsfonts}
\usepackage{graphicx}
\usepackage{hyperref}
\usepackage{multirow}
\usepackage[mathlines]{lineno}
\linenumbers

\usepackage[usenames,dvipsnames]{color}
\usepackage[utf8]{inputenc}
\usepackage{ulem}

\begin{document}

\preprint{Aberdeen - 01/2013}

\title{Analysis of communities in a mythological social network}

\author{P. J. \surname{Miranda}$^1$}
\author{M. S. \surname{Baptista}$^2$}
\author{S. E. \surname{de Souza Pinto}$^{1,2}$}
\email[Corresponding author: ]{desouzapinto@gmail.com}
\affiliation{1.Departamento de F\'isica, Universidade Estadual de Ponta Grossa, 84030-900, Paran\'a, Brazil\\ 2. Institute for Complex Systems and Mathematical Biology, SUPA, University of Aberdeen, Aberdeen, United Kingdom}
 
\date{\today}

\begin{abstract}
{ The intriguing nature of classical Homeric narratives
has always fascinated the occidental culture contributing to philosophy,
history, mythology and straight forwardly to literature. However what would be
so intriguing about Homer's narratives' At a first gaze we shall recognize the
very literal appeal and aesthetic pleasure presented on every page across
Homer's chants in Odyssey and rhapsodies in Iliad. Secondly we may perceive a
biased aspect of its stories contents, varying from real-historical to
fictional-mythological. To encompass this glance, there are some new
archeological finding that supports historicity of some events described
within Iliad, and consequently to Odyssey. Considering these observations and
using complex network theory concepts, we managed to built and analyze a
social network gathered across the classical epic, Odyssey of Homer. Longing
for further understanding, topological quantities were collected in order to
classify its social network qualitatively into real or fictional. It turns out
that most of the found properties belong to real social networks besides
assortativity and giant component's size. In order to test the network's
possibilities to be real, we removed some mythological members that could
imprint a fictional aspect on the network. Carrying on this maneuver the
modified social network resulted on assortative mixing and reduction of the
giant component, as expected for real social networks. Overall we observe that
Odyssey might be an amalgam of fictional elements plus real based human
relations, which corroborates other author's findings for Iliad and
archeological evidences.}
\end{abstract}


\maketitle

\section{Introduction }

The paradigm shift of reductionism to holism are in common use
nowadays, turning the scientist's interests to the interdisciplinary approach
\cite{newmann-book}. This endeavor may be accomplished as far as the fundamental concepts of
complex network theory are applied on problems that may arise from many areas
of study, like social networks, communication, economy, financial market,
computer science, internet, World Wide Web, transportation, electric power
distribution, molecular biology, ecology, neuroscience, linguistics, climate
networks and so on \cite{costa}. As the study of the objects under the network subject
paradigm goes on \cite{newman1}, some classes may arise as function of their general
structure or topology \cite{amaral}. Given these structures, some patterns may be
perceived into the aspect of network measures that, in addiction to one
another, may determine a class of networks \cite{book-socialnetwork}. Considering this one can
define some sort of taxonomy of networks that can be built by the simple
comparison regarding their topological properties \cite{onnela}. With conformity with
this reasoning, a notion of universality also may be used as key-concept to
unify the various kinds of networks into characteristic groups \cite{albert}.

 Based on this concept of universality, Carron and Kenna \cite{carron} proposed
an analysis method to discriminate a given narrative into real or fictional,
based on the social network it would represent. Specifically they analyzed
three classical narratives with uncertain historicity, which was: Beowulf,
Iliad and T\'ain B\'o Cuailnge \cite{carron}. From these was created some sort of social
networks where nodes would represent characters and edges their social
relation in the tale. As result, it generated what was denominated the
mythological networks. This type of network is essentially a social network
with distinguished topological properties. In order to determine what is
distinguished, we must first determine what is common in topological terms.
The real social networks are known to be usually small world \cite{amaral,watts1},
hierarchically organized \cite{ravasz,dorogovtsev,gleiser}, highly clustered, assortatively mixed by
degree \cite{newman2,newman3}, and scale free \cite{amaral,barabasi1}. In addition to these basic
characteristics, real social networks also may show power law dependence of
the degree distribution \cite{amaral}, hold a giant component with less than 90\% of the
total number of vertices, be vulnerable to targeted attacks and hold
robustness to random attacks \cite{albert2}.

 In the other extreme of a social networks configuration derived from
a narrative, there are the fictional social networks. These can be
characterized as being small world, bear hierarchical structure, hold a
exponential dependence of degree distribution, not scale free, size giant
component with more than 90\% of the total vertices, show no assortativity by
degree and being robust to random or targeted attacks. Its properties show
some resembling features of real social networks. However a profound analysis
would show their artificial nature \cite{carron,gleiser,marvel}. Based on these set of
classificatory properties, the authors of the preceding research figured out
that Iliad would be more realistic than fictional in term of its social
network, whilst the other two (Beowulf and T\'ain) would be needed some simple
and reasonable modification to render their network as real \cite{watts1}. Although this
made manipulation, they managed to synthesize a way to analyze folktales,
myths or other classical poems, epics or narratives. This synthesis can be
used to identify sociological structure which is valuable as a tool in the
field of comparative mythology. Conversely it is worth noting and citing

Campbell's work The Hero with a Thousand Faces, where is brought to
light a notion that mythological narratives from diverse culture frequently
share the same structure denominated the monomyth \cite{book-campbell}. With Carron's and
Kenna's results and Campbell's statement, we can build an idea that classic
narratives tend overall to be based on some historicity mixed with some sort
of myth or legend, turning a historical document more attractive to be passed
throughout generations.

Inspired by this preceding speculation, we propose a social network
analysis of the Homeric Greek epic Odyssey \cite{book-odyssey}. Looking for some meaning in
terms of its network topology, we will attempt to classify the resulting net
into real or imaginary as well as considering its implication for the
comparative mythology. In addition to this methodology, we shall also run an
algorithm to discriminate the so called ``communities'' or modules into the
network. The novelty of this work stands is to verify if these sub-social
groupings (i. e. communities) may have meaning, in terms of its characters
composition and within topology, long story shorting: what are their
contribution to the Odyssey's social network? We shall perform this task
through random walk algorithm \cite{pons}. As far as the communities are found, an
interpretation must follow concerning its characters composition, internal
topology and its importance to the whole topological structure may follow.

\section{Network measures}

The first fundamental property that appears with a network is the
total number of vertices N along with the total number of edges between nodes
E. As the net is formed, each node will have a certain number of edges that
make the connection to other vertices; this will be the degree k of the
vertex. The averaging over all degrees gives us the mean degree $\langle k\rangle$ of the network. Exploring a longer bit the degree property, we can
derive $p(k)$, which represents the probability that a given
node has degree $k$, then for most real network the degree distribution holds
for

\begin{equation}
p(k)\sim k^{-\gamma}
\label{eq1}
\end{equation}
for positive and constant 
$\gamma$. 

This is the power law dependency of degree
distribution. For a network this reflects that the nodes are sparsely
connected or there are little nodes with high degree and numerous vertices
with low degree. The scale free characteristic of a network is maintained if
\ref{eq1} 
is satisfied \cite{barabasi1}. Some other important structural properties are also to
be collected in the light of Graph's Theory, likely: the average path length
$l$, the longest geodesic $\ell_{max}$ and the clustering coefficient $C$.
Consider a graph $G$ and its set of vertices $V$. If $d \left( v_i v_j
\right)$ are to be the shortest distance between the vertices $v_i $and $v_j$
, where $v_i$ and $v_j \mathbf{\epsilon}V$ ; assume that $d \left( v_i v_j
\right) = 0$ if $v_j = v_i$ or $v_i $ can't reach $v_j$ . Given these
conditions we can define the average path length $l$ as

\begin{equation}
\ell_{_G} = \frac{1}{n \left( n - 1 \right)}  \sum_{_{i.j}} d \left( v_i v_j
  \right)
\label{eq2}
\end{equation}
where $n$ is the number of vertices of the graph $G$. The longest geodesic,
often known as diameter of a graph, consist simply in the largest value of $d
\left( v_i v_j \right)$ or in other terms, the longest topological separation
between all pairs of vertices of the graph. The third property $C$, quantify
to what extent a given neighborhood of the network is cliqued. If vertices $i$
has $k_i$ neighbors, we find out that the maximal number of potential links
between them will be $k_i \left( k_i - 1 \right) / 2$. Analogous to this we define $n_i$ as the
actual number of links between the $k_i$ neighbors of $i$, the clustering
coefficient  \cite{watts1} of the node shall be defined as,

\begin{equation}
 C_i = \frac{2 n_i}{k_i \left( k_i - 1 \right)} 
\label{eq3}
\end{equation} where the clustering coefficient $C$ for the whole network is simply the
averaging of \ref{eq2}. Many real network show a modular structure implying that
groups of nodes organize in a hierarchical manner into increasingly larger
groups. This feature can be overviewed as the power-law dependency fitting of
the averaged clustering coefficient versus degree \cite{ravasz,dorogovtsev,gleiser}:

\begin{equation}
C \left( k \right) = \frac{1}{k}
\label{eq4}
\end{equation}

Additionally, we test the small-world phenomenon on the network. For
that we sustain that the network will be small world if $\ell \approx
\ell_{{rand}}$ and $C \gg C_{{rand}}$ are both satisfied \cite{watts1}.

 Where $\ell_{{rand}}$ and $C_{{rand}}$, are respectively, the
average path length and the clustering coefficient for a random network built
with the same size $\left( N \right)$ and degree distribution \cite{watts1}. We also
intent to measure the assortative mixing by degree, which brings the notion
that nodes of high degree often associate with similarly highly connected
nodes, while nodes with low degree associate with other less linked nodes.
This quantity is given by the simple Pearson correlation r for all pairs of
$N$ nodes of the network.  Newman showed that real social network tend to be
assortatively mixed by degree, conversely Gleiser sustained that
disassortativity of social network may signal artificiality, and in our
context, a fictional social network \cite{gleiser,newman2}.

The size of the giant component $G_c$ is an important network property
which, in some fashion, measure the connectivity capturing the maximal
connected components of a network \cite{albert}. It is also stated that in scale free
networks, removal of influential nodes causes the giant component to break
down quickly demonstrating vulnerability. This is an important feature of real
social network may have. However the process depends on how we define the
importance of a node in the network. As well as degree, the ``betweenness''
centrality of a given node $g_{\ell}$ indicates how influential that node is in the
net. This measure can be defined as an amalgamation of the degree and the
total number of geodesics that pass through a vertex \cite{freeman}. If $\sigma \left(
i, j \right)$ \ is the number of geodesics between vertices $i$ and $j$ and
number of these which pass through node $\ell$ is $\sigma_1 \left( i, j \right)$
, then the betweenness centrality of $\ell$ \ shall be given by

\begin{equation}
g_{\ell} = \frac{2}{\left( N - 1 \right) \left( N - 2 \right)} \sum_{i \neq j} 
  \frac{\sigma_{\ell} \left( i, j \right)}{\sigma \left( i, j \right)}
\label{eq5}
\end{equation} $g_{\ell}$ will be 1 if all geodesics pass through $\ell$. With this node's importance
defined, it is possible to perform the target attack, which is the removal of
the most important nodes seeing how the size of the giant component behaves
after the removal. In addition to the target attack, we shall realize a random
attack where, differently from the target attack; the vertices to be removed
are chosen at random. The main difference between these two kinds of attacks
may show us some kind of intrinsic organization within the social network
\cite{albert2}.

As a complement of all topological measurements described until now, we
also applied an algorithm called Walktrap that captures the dense subgroups
within the network often known as ``communities'' via random walk \cite{pons}. This
method allows us to describe the ``communities'' composition in terms of its
characters array and topological configurations.

\section{Odyssey's poem analysis}

Along with Iliad, the Odyssey of Homer express with fierce and beauty
the wonders of the remote Greek civilization. The epics date around the VIII
century B.C., after the writing system development using the Phoenician
alphabet \cite{book-odyssey2,intro-odyssey}. It is also known that Odyssey carry some echoes from the
Trojan War narrated mainly on Iliad. Recalling again Carron's and Kenna's
paper, from their three myths analyzed, the network of characters from Iliad
showed properties most similar to those of the real social networks. In
addiction they maintained that this similarity perhaps reflects the
archeological evidence supporting the historicity of some conflict occurred
during the XII century B.C. \cite{archeology,kraft}. The poem's title (Odyssey) comes from
the name of the protagonist, Odysseus (or Ulysses, for the roman adaptation),
son and successor of Laertes, King of Ithaca and husband of Penelope. The epic
has its center scenario on the protagonist journey back home after his
participation on the Trojan War. This saga takes Odysseus ten years to reach
Ithaca after the ten years of warring \cite{book-odyssey}. The epic poem is composed by 24
chants in hexameter verses, where the tale begin 10 years after the War in
which Odysseus fought siding with the Greeks. Worth noting that the narrative
has inverse order: it starts with the closure, or the Assembly of the Gods
when Zeus decides Odysseus's journey back to home. The text is structured on
four main parts: the first (chants I to IV), entitled ``Assembly of the
gods''; the second (chants V to VIII), ``The new assembly of the gods''; the
third (chants IX to XII), ``Odysseus's Narrative''; and forth (chants XIII to
XXIV), ``Journey back home''.

 Odyssey's masterwork after all, holds a set of adventures often
considered more complex than Iliad; it has many epopee aspects that are close
to human nature, while the predominant aspect of Iliad is to be heroic,
legendary and of godlike wonders. However there is a consensus that Odyssey
completes Iliad picture of the Greek civilization, and together they hold the
very witness geniality of Homer, being both pieces of fundamental importance
to universal poesy in the occident \cite{book-odyssey}.

\begin{figure}[htbp!]
\center
\includegraphics[width=0.7\textwidth]{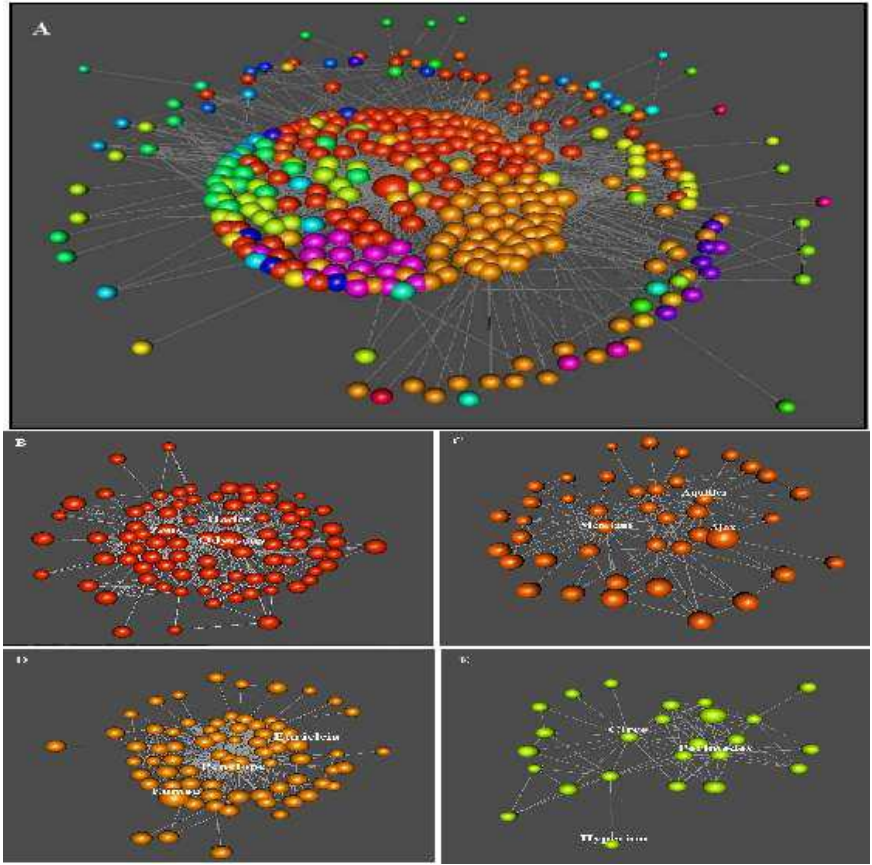}
\caption{(A) Represents Odyssey's social network, the coloring of the vertex determine to which community they belong. The vertex size is based on its importance on the network; (B), (C) and (D) show some of the most important communities to the whole network.}
\label{fig1}
\end{figure}

 As a careful textual analysis was
performed, we managed to identify 342 unique characters bounded socially by
1747 relations \ref{fig1}. We should point out that this network may be
socially limited; it rather captures some spotlights on the societies from
that time \cite{albert,carron}. We define the social relation between two character when
they've met in the story, speak directly to each other, cite one another to a
third or when it is clear they know each other. To avoid some possible
misleading interpretation of the poem's social relations, we studied different
translations and editions of Odyssey \cite{book-odyssey,book-odyssey2,intro-odyssey}. The basic differences
from the Odyssey's translations generated no significant deviation as the
network creation process was made.

\section{Network topology and analysis results}

\begin{figure}[htbp!]
\center
\includegraphics[width=0.7\textwidth]{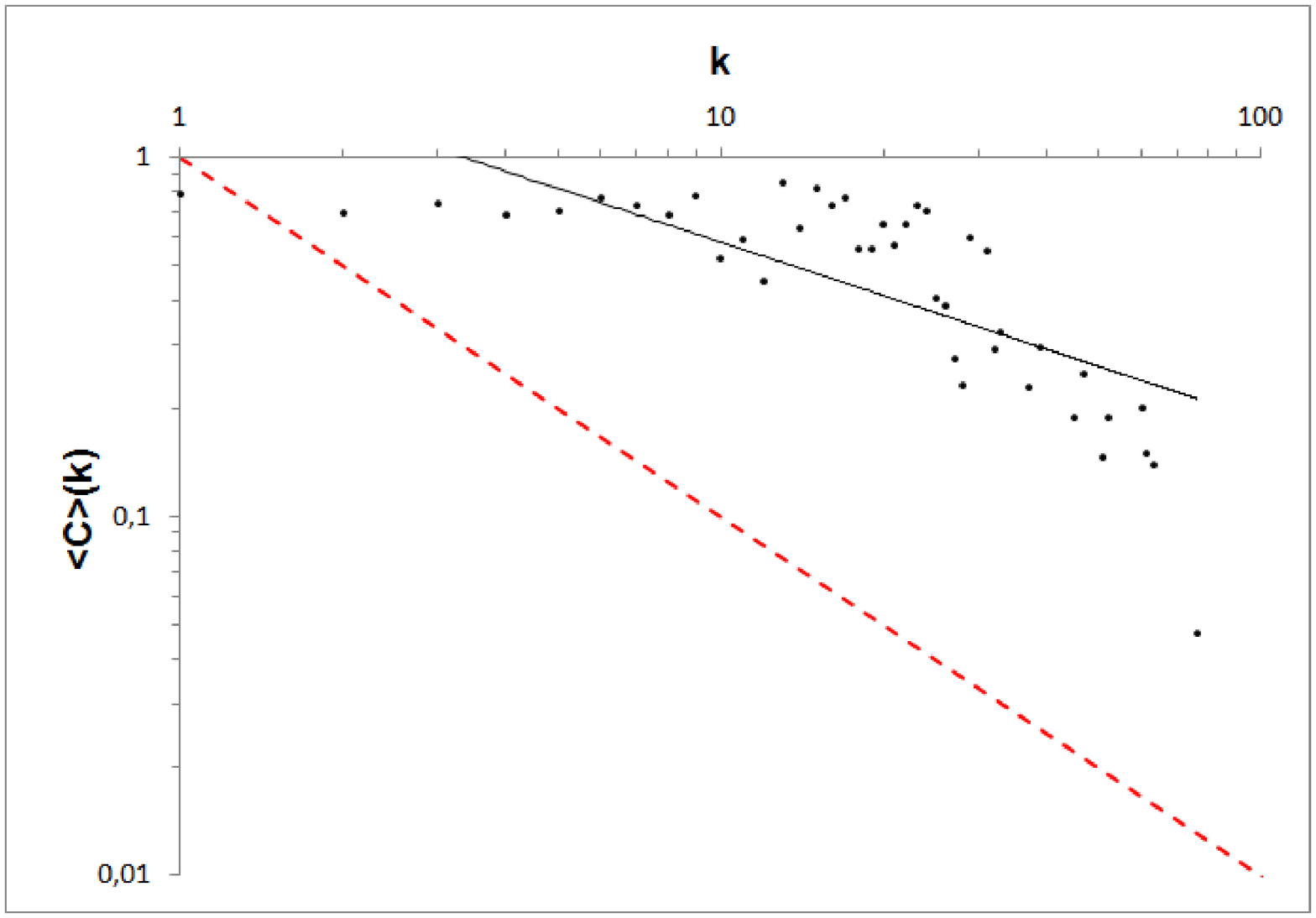}
\caption{Mean clustering coefficient per degree with the red dashed line holding for the power law $1/k$.}
\label{fig2}
\end{figure}

 A summary (Table \ref{tab1}) of the found topological properties was compiled
along with Carron's and Kenna's results for their described mythological
networks \cite{watts1}. As expected, the social network analyzed has average path length
similar to a random network build with same size and average degree.
Additionally it also has high clustering coefficient compared to random
network indicating the small world phenomenon. The hierarchical feature of the
network is displayed on \ref{fig2}, where the mean clustering coefficient per
degree is plotted $\left( C \left( k \right) \times k \right)$. It is
possible to verify that nodes with smaller degree present higher clustering
than those with higher degree, the decay these relation may follow
approximately \ref{eq3}. We may interpret that high degree vertices integrate the
small communities, generating the unification of the whole network.

\begin{table}
\begin{tabular}{|c|c|c|c|c|c|c|c|c|c|c|}
\hline
{\bf Network} & $\mathbf{N}$ & $\mathbf{E}$ & $\mathbf{\langle k \rangle}$ & $\mathbf{\ell}$ & $\mathbf{\ell_{rand}}$ & $\mathbf{\ell_{max}}$ & $\mathbf{C}$ & $\mathbf{C_{rand}}$ & $\mathbf{G_C}$ & $\mathbf{r}$\\
\hline
Odyssey & 342 & 1747 & 10.22 & 2.58 & 2.47 & 6 & 0.27 & 0.11 & 342(100\%)  & -0.15\\
\hline
Odyssey* & 318 & 1129 & 7.10 & 4.08 & 3.10 & 11 & 0.54 & 0.06 & 274(86\%) & 0.09\\
\hline
Iliad & 716 & - & 7.40 & 3.54 & 3.28 & 11 & 0.57 & 0.01 & 707(98.7\%) & -0.09\\
\hline
Beowulf & 74 & - & 4.45 & 2.37 & 2.88 & 6 & 0.69 & 0.06 & 50(67.5\%) & -0.10 \\
\hline
T\'ain & 404 & - & 2.76 & 2.76 & 3.32 & 7 & 0.82 & 0.02 & 398(98.5\%) & -0.33\\
\hline
Beowulf* & 67 & - & 3.49 & 2.83 & 3.36 & 7 & 0.68 & 0.05 & 43(64.2\%) & 0.01\\
\hline
T\'ain* & 324 & - & 3.71 & 3.88 & 4.41 & 8 & 0.69 & 0.01 & 201(62\%) & 0.04\\
\hline
\end{tabular}
\caption{Size ($N$), number of edges ($E$), average path length ($\ell$), diameter ($\ell_{max}$), clustering ($C$), size of giant component ($G_c$ and assortitativity ($r$). Odyssey*, Beowulf* and Táin are teh same original network with some character modification.}
\label{tab1}
\end{table}

\begin{figure}[htbp!]
\center
\includegraphics[width=0.7\textwidth]{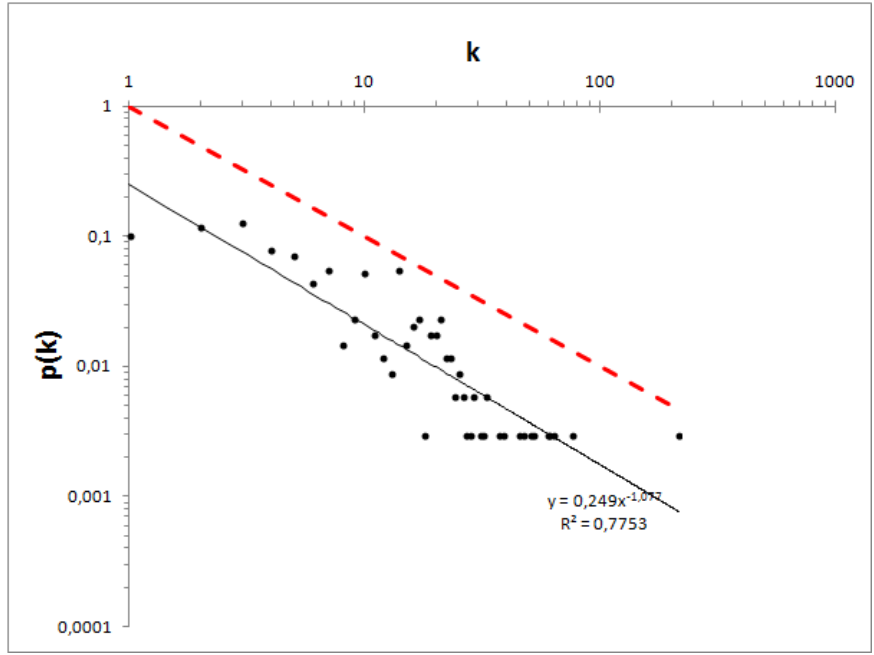}
\caption{Degree distribution of Odysseys1s network with the red dashed line representing the power law distribution.}
\label{fig3}
\end{figure}

 The observed giant component of Odyssey contains all the vertices of
the network, suggesting two possible reasoning: the giant component phenomenon
didn't actually occur or the data set is limited to its appearance. Following
with that thought, the directed attack showed that the network's topology
depends mostly on central characters, considering that its giant component is
little affected (Table \ref{tab2}). However if we remove, at the same proportion,
nodes chosen independently from their ``betweeness'' centrality, we see a
certain robustness of the giant component size. Vulnerability to targeted
attack and resilience to random attacks indicate that studied network may be
scale free. Analogous to this, we show in the Figure \ref{fig3} the degree distribution
and its power dependence for $Y = 1, 2 \pm 1$ (with $X^2 d f = 0, 06$),
demonstrating that the network is actually free scale.

\begin{table}
\begin{tabular}{|l|l||l|l|}
\hline
 \multicolumn{2}{|c||}{$\mathbf{G_c}$ {\bf for a targeted attack}}&\multicolumn{2}{|c|}{$\mathbf{G_c}$ {\bf for a random attack}}\\
\cline{1-4}
\hline\hline
No attack & 3442 (100\%)& No attack & 342 (100\%)\\
\hline
5\% & 274 (79.6\%) & 5\% & 332 (93.6\%)\\\hline
10\% & 188 (54.6\%) & 10 \% & 309 (89.9\%)\\\hline
15\% & 163 (47.3\%) & 15\% & 282 (81.9\%) \\\hline
20\% & 121 (35.1\%) & 20\% & 273 (79.3\%)\\\hline
25\% & 41 (11.9\%) & 25 \% & 248 (72\%)\\
\hline
\end{tabular}
\caption{Directed and random attacks along with size of giant component $G_c$ response in terms of absolute and relative number  of nodes.}
\label{tab2}
\end{table}

Early we stated that real social networks tend to be assortatively
mixed by degree while disassortativity of social network signal artificiality.
As result the network is found to be disassortatively. However this
disassortativity may also reflect conflictual nature into the epic. The same
result were observed on Iliad and T\'ain, where main characters find
themselves on confrontation with enemies which seems to have any other
relations to the rest of the tale, resulting on a high number of loose
vertices connected to central vertices. These phenomena should drive the
overall degree assortativity to negative values \cite{carron}. This may also be due to
high linked central characters be connected to low accessory characters
considering the epic's context. To overcome this problem we removed those
nodes that carry the mythological background into the story, verifying its
effect on the size of the giant component and degree assortativity. These
characters would be those with central importance for the epic known as main
heroes, legends, godlike or even gods. We tested the removal of high
importance in terms of topology and epics continuance, the results are resumed
on Table \ref{tab3}. It is notable that as the removing of central characters
follows, the network tends to demonstrate a real social network pattern. It
becomes assortatively mixed by degree and bears size of the giant component
less than 90\%. Besides giant component and assortativity we must analyze the
further topological modification. The resulting modified network is named
Odyssey* compiled on Table \ref{tab1}. As result, naturally we expected the decay on
the observed number of vertices and edges. Recalling that the removal was for
the most important nodes the mean degree felt as well. Unfortunately the
difference between the average path length and the random network average path
length increased, so the network loses some of its small world pattern,
however the clustering coefficient difference is enhanced. As this follow, we
still consider the network to be small-world. Another interesting fact is the
resembling of the mean degree and diameter of Iliad and Odyssey*. Due to
modifications, the analyzed epic becomes closer to Iliad, something worth
noting since Iliad was rendered real based \cite{carron}. Additionally the degree
distribution showed no significant difference after the modification.

The previous manipulation implies that the network can be perceived
as an amalgam of fictional and real aspects. Which the fictional effects,
built up by fictional characters (heroes, gods and monsters) form within
themselves a back-bone effect to the built of the social network to which all
the other relation shall connect. We verify this network's dependence on the
next section more explicitly. Overall we may say that these myths all over the
tale become intertwined with real backgrounds which are formed by humans
relations, tribes, families or some other sort of human cliques. This brings
us back to the monomyth concept of Campbell, which most of the myths have
common structures based on a main character and his crusade, being the
secondary elements of the crusade based or not on sociological relation in the
epoch \cite{book-campbell}. Considering this and our given findings, it is reasonable to think
that Homer himself could've intended to write a myth mixed with historical
facts in order to increase its appeal hoping for the next generations to reach
it \cite{intro-odyssey,book-odyssey3}.

\section{Network's communities' composition}
\begin{figure}[htbp!]
\center
\includegraphics[width=0.7\textwidth]{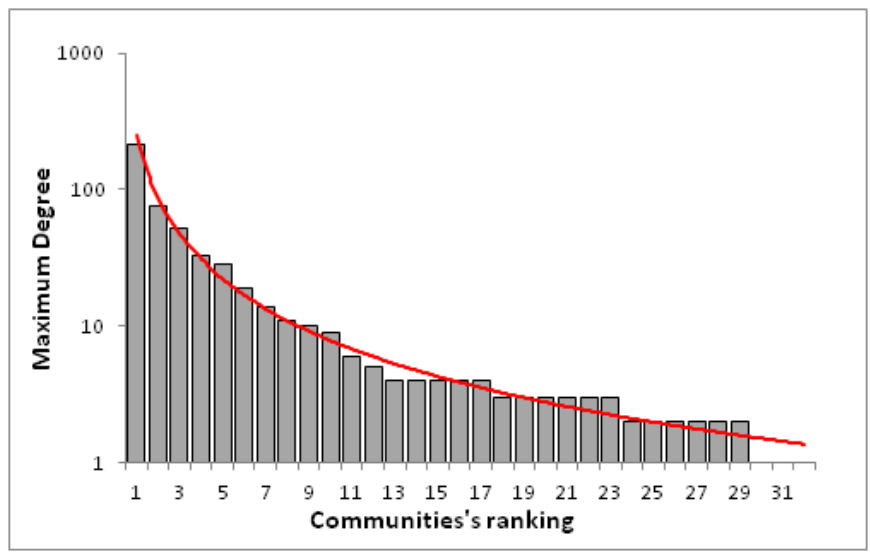}
\caption{Maximum degree of each community plotted on ranking.}
\label{fig4}
\end{figure}

\begin{figure}[htbp!]
\center
\includegraphics[width=0.7\textwidth]{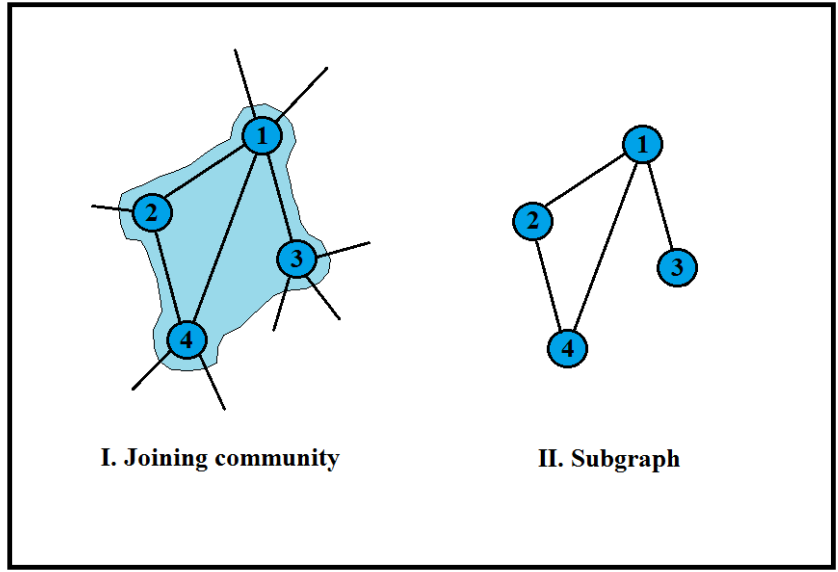}
\caption{(I) Joining Community: the nodes keep their degree and topological dependences with the rest of the network. (II) Subgraph: the topological quantities depend only on the community alone.}
\label{fig5}
\end{figure}

Through the walktrap algorithm we could identify 32 communities that
compound Odyssey's social network, the max degree distribution, as a community
importance ranking, can be found on Figure \ref{fig4}. Additionally the most
influential communities $\left( n \geq 10 \right)$ \ and their topological
measures can be found on table IV, comprising 10 communities. The features
described was the size $\left( N \right)$, maximum degree $\left( k_{\max}
\right)$, mean degree $\left\langle k \right\rangle$, average path length
$\left( \ell \right)$, average path length for a randomly created community \
$\left( \ell_{{rand}} \right)$, clustering coefficient \ $\left( C \right)$,
clustering coefficient for a randomly created community $C_{{rand}}$,
assortativity $\left( r \right)$, small world ($\ell \approx \ell_{{rand}}$ and
$C \gg C_{{rand}}$), degree distribution fitting and hierarchy test. As
far as the concept of community is difficult to clearly define, we choose to
recognize them using two different "views" for this work. Although we may
perceive the communities as part of the complete network, that we call joining
communities, we can also discriminate them separately from the network, which
represents the subgraphs (see Figure \ref{fig5}). This differentiation seems to be
trivial but is actually very important to the topological measures within
communities. This happens because topological properties of the communities
shall vary as long as if they belong or not the whole network, except for the
number of vertices, N.

\begin{table}
\begin{tabular}{|l|c|c|}
\hline
{\bf Character} & {\bf Assortativity} & {\bf Size of giant component}\\
\hline
\hline
Complete network & -0.15 & 100\%\\
\hline
Odysseus removal & -0.07 & 97\%\\
\hline
plus Zeus removal & -0.06 & 97\%\\
\hline
plus Telemacus removal & -0.03 & 95\%\\
\hline
plus Athena removal & -0.04 & 93\%\\
\hline
plus Penelope removal & -0.04 & 92\%\\
\hline
plus Menelaus removal & 0.007 & 92\%\\
\hline
plus Hades removal & 0.03 & 92\%\\
\hline
plus Poseidon removal & 0.06 & 91\%\\
\hline
plus Persephone removal & 0.09 & 86\%\\
\hline
\end{tabular}
\caption{Odyssey's network main characters removal along with assortativity and giant component responses.}
\label{tab3}
\end{table}

On a first look at Table \ref{tab4}, we realize that community A have a
especial tendency of organization when compared to others community. The
property that points out this effect is the higher mean degree as the
community is took away from the network; this incites us to think that this
part can sustain itself as a network. The small world and the scale free
phenomena also have been observed on this community, confirming its global
function as a social network. Yet, this network doesn't seem to have a
hierarchical structure, so hierarchical feature of the total network must
arise as the several communities combine themselves. The number of nodes of
each community, that follows a power law distribution, corroborates to this
behavior, suggesting that the most relevant communities combine themselves
with the lesser ones choosing the most connected elements to make links
between communities. We also maintain that community $A$ is special; actually it
acts as a back-bone spine to the other communities for form the total network.

Well, we must now consider the character composition of each of the
most important community. The process of random walk algorithm discriminates
the communities choosing each element to belong or not to a dense set, but to
this work it also managed to separate each community on terms of the epic's
chants character composition. For example, the community A contain all the
elements of the chants concerning the ``Assembly of the gods''; the community
$B$ is composed by the most remarkable heroes that played an important role on
the Troy's War; community $C$ synthesizes the elements present on the episodes
that narrates the events occurring on Ithaca while Odysseus's Journeys, most
of them comprising Penelope's servants and suitors; community $D$ captures some
secondary characters like nymphs, godlike monsters and minor gods, they come
along because there is some secondary stories that enclose them into the main
epic; community $E$ takes account the sons and daughters of Nestor, capturing
the episode where Telemacus search for his father; community $F$ is very
important, it ensembles several epics scenarios of Odysseus's Journey,
capturing the passage to the Lotofagus's Island, Circe's Lair, Hiperyon Sun's
Island and Sila and Caribdis episode. This community also contains most of the
main journey's comrades of Odysseus; community G represents the Odysseus
passage to the land of Phaeacians, containing all elements of that episode;
and finally community H captures the Odysseus passage to Eolus's Island, so
its characters are composed mainly by his sons and daughters. It is probable
that the used algorithm separated the Odyssey's network into the societies the
tale itself narrates.



\section{Conclusions}

 Our analysis demonstrated that Odyssey's social network is small world,
highly clustered, slightly hierarchical and resilient to random attacks. This
configuration is known to belong for most real social networks. However and
additionally we also found that it is vulnerable to target attack, hold for
power law dependence of degree distribution and is scale free. Although
further modification were made to the network, the assortativity degree and
non-total giant component was confirmed; these was achieved when high
connected nodes were removed, mainly those representing heroes, gods and
legends. This procedure lead us to think that mythical elements on Odyssey's
epic drag the network to a fictional aspect, ergo its fundamental background
is purely real considering its social topological terms. This set of observed
results; both modification and interpretation lead us to conclude that Odyssey
may be fruit of a mixture of myth and historical based societies that
corroborate with Carron and Kenna finding of Iliad and the archeological
evidence that supports its events.

\begin{table}
\begin{tabular}{|c|c|c|c|c|c|c|c|c|c|c|c|c|}
\hline
\hline
Community & Type & $N$ & $k_{max}$ & $\langle k \rangle$ & $\ell$ & $\ell_{rand}$ & $C$ & $C_{rand}$ & $r$ & Small World & $p(k)$ & Hierarchy\\
\hline
\hline
\multirow{2}{*}{Com. A} & joining & 83 & 214 & 8.30 & 2.08 & 2.41 & 0.57 & 0.23 & * & no & $NA$ & no\\
& subgraph & 83 & 66 & 12.8 & 2.09 & 2.09 & 0.62 & 0.27 & -0.29 & yes & power law & no\\
\hline
\multirow{2}{*}{Com. B} & joining & 42 & 52 & 9.10 & 2.02 & 2.09 & 0.67 & 0.26 & * & no & $NA$ & no\\
& subgraph & 42 & 29 & 6.10 & 2.09 & 2.02 & 0.64 & 0.42 & -0.24 & no & $NA$ & no\\
\hline
\multirow{2}{*}{Com. C} & joining & 73 & 76 & 13.40 & 2.05 & 2.30 & 0.61 & 0.22 & * & no & $NA$ & no\\
& subgraph & 73 & 46 & 11.00 & 2.12 & 2.00 & 0.59 & 0.54 & -0.28 & no & $NA$ & no\\
\hline
\multirow{2}{*}{Com. D} & joining & 10 & 10 & 3.90 & 1.86 & 2.64 & 0.66 & 0.19 & * & no & $NA$ & no\\
 & subgraph & 10 & 6 & 2.20 & 2.00 & 1.94 & 0.28 & 0.34 & -0.40 & no & $NA$ & no\\
\hline
\multirow{2}{*}{Com. E} & joining & 11 & 19 & 9.81 & 1.30 & 1.85 & 0.65 & 0.26 & * & no & $NA$ & no\\
& subgraph & 11 & 9 & 6.18 & 1.30 & 1.30 & 0.70 & 0.70 & -0.30 & no & $NA$ & no\\
\hline
\multirow{2}{*}{Com. F} & joining & 25 & 28 & 9.60 & 1.78 & 2.19 & 0.66 & 0.33 & * & no & $NA$ & no\\
& subgraph & 25 & 19 & 6.88 & 1.82 & 1.77 & 0.66 & 0.55 & -0.26 & no & $NA$ & no\\
\hline
\multirow{2}{*}{Com. G} & joining & 20 & 33 & 14.9 & 1.31 & 2.00 & 0.74 & 0.18 & * & no & $NA$ & no \\
& subgraph & 20 & 16 & 12.5 & 1.31 & 1.31 & 0.82 & 0.82 & -0.12 & no & $NA$ & no\\
\hline
\multirow{2}{*}{Com. H} & joining & 13 & 14 & 14.00 & 0.92 & 1.88 & 1.00 & 0.20 & * & no & $NA$ & no\\
& subgraph & 13 & 12 & 12.00 & 0.92 & 0.92 & 1.00 & 1.00 & -0.01 & no & $NA$ & no\\
\hline
\end{tabular}
\caption{Type (joining community or sub graph), size ($N$), maximum degree ($k_{max}$), mean degree $\langle k \rangle$, average path length ($\ell$), average path length for a randomly created community  ($\ell_{rand}$), clustering coefficient ($C$), clustering coefficient for a randomly created community ($C_{rand}$), assortativity ($r$), small world ($\ell \approx \ell_{rand}$ and $C >> C_{rand} $), degree distribution and hierarchy. OBS.: $NA$ holds for any fit.}
\label{tab4}
\end{table}

This work has been made possible thanks to the partial financial support from the following Brazilian research agencies: CNPq, CAPES and Funda\c c\~ao Arauc\'aria.




\begin{thebibliography}{99}
\bibitem{newmann-book} Newman, M. E. J. Networks, An Introduction. (Oxford University Press, New York) 2010.


\bibitem{costa}Costa, Luciano da Fontoura; Oliveira, Osvaldo N., Jr.; Travieso, Gonzalo; {\it et al}.  Analyzing and modeling real-world phenomena with complex networks: a survey of applications. Advances in Physics {\bf 60}, 329 (2011).

\bibitem{newman1} Newman, M. E. The Structure and function of complex networks, SIAM Rev., 45 (2003) 167.

\bibitem{amaral} Amaral, L. A., Scala, A., Barth\'el\'emy, M. and Stanley, H. E. Classes of small-world networks. Proc. Natl. Acad. Sci. U.S.A., 97 (2000) 11149.

\bibitem{book-socialnetwork}  Wasserman, F. and Faust, K. Social Network Analysis. (Cambridge University Press, Cambridge) 1994.

\bibitem{onnela} Onnela, Jukka-Pekka; Fenn, Daniel J.; Reid, Stephen; {\it et al.} Taxonomies of networks from community structure. Phys. Rev. E {\bf 86}, 036104 (2012).

\bibitem{albert} Albert, R. and Barab\'asi, A.-L. Statistical mechanics of complex networks. Rev. Mod. Phys., 74 (2002) 47.

\bibitem{carron} Carron, P.M. and Kenna, R. Universal properties of mythological networks. E.P.L., 99 (2012) 28002.

\bibitem{watts1} Watts, D. J. and Strogatz, S. H. Collective dynamics of small-world networks. Nature, 393 (1998) 440442.

\bibitem{ravasz} Ravasz, E. and Barab\'asi, A.-L. Hierarchical organization on complex
networks. Phys. Rev. E., 67 (2003) 026112.

\bibitem{dorogovtsev} Dorogovtsev, S. N., Goltsev, A. V. and Mendes, J. F. F. Pseudofractal
scale-free web. Phys. Rev. E., 65 (2002) 066122.

\bibitem{gleiser} Gleiser, P. M. How to become a superhero. J. Stat. Mech. (2007) P09020.


\bibitem{newman2} Newman, M. E. J. Assortative mixing in networks. Phys. Rev. Lett., 89
(2002) 208701.

\bibitem{newman3} Newman, M. E. J. and Park, J. Why social networks are different from
other types of networks. Phys. Rev. E., 68 (2003) 036122.

\bibitem{barabasi1} Barab\'asi, A.-L. and Albert, R. Emergence of Scaling in random networks.
Science, 286 (1999) 509512.

\bibitem{albert2} Albert R., Jeong, H. and Barab\'asi, A.-L. Error and attack tolerance of
complex networks. Nature, 406 (2000) 378.

\bibitem{marvel} Alberich, R., Miro-Julia, J. and Rosell\'o, F. Marvel Universe looks
almost like a real social network. Cond-mat/0202174.

\bibitem{book-campbell} Campbell, J. The Hero with a Thousand Faces. (Princeton University Press,
Princeton) 1949.

\bibitem{book-odyssey}  Odyssey of Homer, translated by Mendes, M. Edited by Rodrigues, A. M.
(Editora da Universidade de S\~ao Paulo) 2000.

\bibitem{pons} Pons, P. and Latapy, M. Computing communities in large networks using
random walks. Computer and Information Sciences {\bf 3733}, 284 (2005).


\bibitem{freeman} Freeman, L. C. A set of Measures of Centrality Based on Betweenness.
Sociometry, 40 (1977) 35.

\bibitem{book-odyssey2} The Odyssey, translated by Rieu, E. V. (Penguin Classics, London) 1946.

\bibitem{intro-odyssey}  Introduction to The Odyssey, Rieu, E. V. (Penguin Classics, London) 2003.

\bibitem{archeology} Korfman, M. Archeology, 57 (2004) 36.

\bibitem{kraft} Kraft, J. C., Rapp, G., Kayan, I. and Luce, J. V. Geology, 31 (2003) 163.

\bibitem{book-odyssey3} Odyssey of Homer, translated by the Priests Palmeira, E. D. and Correia,
M. A. (Livraria S\'a da Costa Editora, Portugal) 1994.

 
\end{thebibliography}
\end{document}